\documentclass[preprint,showpacs,preprintnumbers,amsmath,amssymb]{revtex4}
\usepackage{graphicx}
\usepackage{dcolumn}
\usepackage{bm}

\newcommand\br{\begin{eqnarray}}
\newcommand\er{\end{eqnarray}}
\newcommand\be{\begin{equation}}
\newcommand\ee{\end{equation}}

\newcommand\bc{\begin{center}}
\newcommand\ec{\end{center}}

\newcommand\PRL[3]{\textsl{Phys. Rev. Lett.} \textbf{#1}, #3 (#2)}

\newcommand\PRD[3]{\textsl{Phys. Rev.} \textbf{D#1}, #3 (#2)}

\newcommand\PLB[3]{\textsl{Phys. Lett.} \textbf{#1B}, #3 (#2)}
\newcommand\CQG[3]{\textsl{Class. Quantum Grav.} \textbf{#1}, #3 (#2)}

\newcommand\AoP[3]{\textsl{Ann. of Phys.} \textbf{#1}, #3 (#2)}

\newcommand\IJMPA[3]{\textsl{Int. J. Mod. Phys.} \textbf{A#1}, #3 (#2)}

\newcommand\JPA[3]{\textsl{J. Physics} \textbf{A#1}, #3 (#2)}

\newcommand\MPLA[3]{\textsl{Mod. Phys. Lett.} \textbf{A#1}, #3 (#2)}

\begin{document}


\title{Emerging Universe from Scale Invariance}

\author{Sergio del Campo}
\email{sdelcamp@ucv.cl} \affiliation{ Instituto de F\'{\i}sica,
Pontificia Universidad Cat\'{o}lica de Valpara\'{\i}so, Avenida
Brasil 2950, Casilla 4059, Valpara\'{\i}so, Chile.}
\author{Eduardo I. Guendelman}
\email{guendel@bgu.ac.il} \affiliation{ Physics Department, Ben
Gurion University of the Negev, Beer Sheva 84105, Israel}
\author{Ram\'on Herrera}
\email{ramon.herrera@ucv.cl} \affiliation{ Instituto de
F\'{\i}sica, Pontificia Universidad Cat\'{o}lica de
Valpara\'{\i}so ,  Avenida Brasil 2950, Casilla 4059,
Valpara\'{\i}so, Chile.}

\author{Pedro Labra$\tilde{n}$a}
\email{plabrana@ubiobio.cl} \affiliation{ Departamento de
F\'{\i}sica, Universidad  del B\'{\i}o B\'{\i}o, Avenida Collao
1202, Casilla 5-C, Concepci\'on, Chile.}

\bigskip

\begin{abstract}

 We consider  a scale invariant model which includes a $R^{2}$ term in
action and show that a stable "emerging universe" scenario is
possible. The model belongs to the general class of  theories,
where an integration measure independent of the metric is
introduced. To implement scale invariance (S.I.), a dilaton  field
is introduced. The integration of the equations of motion
associated with the new measure gives rise to the spontaneous
symmetry breaking (S.S.B) of S.I. After S.S.B. of S.I. in the
model with the $R^{2}$ term (and first order formalism applied),
it is found that a non trivial potential for the dilaton is
generated. The dynamics of the scalar field becomes non linear and
these non linearities are instrumental in the stability of some of
the emerging universe solutions, which exists for a parameter
range of the theory.

\end{abstract}


\maketitle
\section{Introduction}
In modern cosmology our notions concerning the early universe have introduced
a new element, the inflationary phase of the early universe \cite{Inflation}, which provides
an attractive scenario for solving
 some of the fundamental puzzles of the standard Big Bang model,
like the horizon and the flatness problems as well as providing a
framework for sensible calculations of primordial density perturbations.

Even in the context of the inflationary scenario however one
encounters the initial singularity problem which remains unsolved,
showing that the universe necessarily had a beginning for generic
inflationary cosmological model\cite{singularities}.

Recently models that can avoid those conclusions have been
discovered \cite{emerging1,emerging2,emerging3,emerging4,
emerging5,emerging6,emerging7}. The way to escape the singularity
in these models is to violate the geometrical assumptions of these
theorems, which assume i) that the universe has open space
sections ii) the Hubble expansion is always greater than zero in
the past. In \cite{emerging1},\cite{emerging2} the open space
section condition is violated since closed Robertson Walker
universes with $k=1$ are considered and the Hubble expansion can
become zero, so that both i) and ii) are avoided.

In \cite{emerging1} even models based on standard General Relativity, ordinary matter and minimally coupled scalar fields
were considered and can provide indeed a non singular (geodesically complete) inflationary universe, with a past eternal Einstein
static Universe that eventually evolves into an inflationary Universe.

Those most simple models suffer however from instabilities, associated with the instability of the Einstein static universe.
The instability is possible to cure by going away from  GR, considering non perturbative corrections to the Einstein`s field
equations in the context of the loop quantum gravity\cite{emerging3}, a brane world cosmology with a time like extra dimension\cite{emerging4}, considering the
Starobinski model for radiative corrections (which cannot be derived from an effective action)\cite{emerging5}, exotic matter\cite{emerging6} or $f(R)$ theories in the presence of perfect fluids with $w<0$ \cite{emerging7}. In addition to this, the consideration of a Jordan Brans Dicke model also can provide
a stable initial state for the emerging universe scenario \cite{emerging8}.

In this paper we propose a different theoretical framework where
such emerging universe scenario is realized in a natural way,
where instabilities are avoided and a successful inflationary
phase with a graceful exit can be achieved.

We work in the context of a two measures theory (TMT) \cite{TMT1}
and more specifically in the context of the scale invariant
realization of such theories \cite{TMT2, TMT3, TMT4,TMT5}. These
theories can provide a new approach to the cosmological constant
problem and can be generalized to obtain also a theory with a
dynamical spacetime \cite{dyn} . We will show how the stated goals
can be achieved in the framework of the TMT models.

The paper will be organized as follows: In section II the
effective field Eqs.  in the Einstein frame, give rise to an
effective potential for a dilaton field (needed to implement a
model with global scale invariance) which presents a flat region.
This model can describe an emerging universe scenario. In section
III, we look at the generalization of this model \cite{TMT5} by
adding a curvature square, "$R^{2}$", and show that the resulting
model contain a flat region. Furthermore the universe can evolve
into an inflationary state which can undergo a successful graceful
exit. We end with a discussion and conclusions section.

\section{ The Emerging Scenario after the introduction of a $R^{2}$ term}

It is possible to obtain a model that through a spontaneous
breaking of scale invariance can give us an emerging universe
scenario. We consider  the scale invariant action of the form
\cite{TMT1,TMT2}

\begin{equation}
S_{L} =  \int L_{1} \Phi d^{4} x  +  \int L_{2} \sqrt{-g}   d^{4}
x + \epsilon  \int (g^{\mu\nu} R_{\mu\nu} (\Gamma))^{2} \sqrt{-g}
d^{4} x.
\end{equation}
Here, $L_1$ and $L_2$ are given by
\begin{equation}
L_{1} = \frac{-1}{\kappa} R(\Gamma, g) + \frac{1}{2} g^{\mu\nu}
\partial_{\mu} \phi \partial_{\nu} \phi - V(\phi),
\end{equation}
and
\begin{equation}
L_{2} = U(\phi),
\end{equation}
respectively. Also, $\Phi$ is defined as $\Phi =
\varepsilon^{\mu\nu\alpha\beta}  \varepsilon_{abcd}
\partial_{\mu} \varphi_{a} \partial_{\nu} \varphi_{b} \partial_{\alpha}
\varphi_{c} \partial_{\beta} \varphi_{d}$ where $\varphi_{a} \,(a
= 1,2,3,4)$ is a 4-scalar fields. The last term in the previous
action  is not only globally scale invariant, but also locally
scale invariant, that is conformally invariant.

Let us see which are the equations of motion. The variation of the
action respect to the 4-scalar fields, $\varphi_a$, yields to
\begin{equation}
A^{\mu}_{a} \partial_{\mu} L_{1} = 0,
\end{equation}
where  $A^{\mu}_{a} = \varepsilon^{\mu\nu\alpha\beta}
\varepsilon_{abcd} \partial_{\nu} \varphi_{b} \partial_{\alpha}
\varphi_{c} \partial_{\beta} \varphi_{d}$, and since $det
(A^{\mu}_{a}) =\frac{4^{-4}}{4!} \Phi^{3}$ we get $\partial_{\mu}
L_{1} = 0$, for $\Phi\neq 0$, and thus $L_1$ becomes
\begin{equation}
L_{1} = \frac{-1}{\kappa} R(\Gamma,g) + \frac{1}{2} g^{\mu\nu}
\partial_{\mu} \phi \partial_{\nu} \phi - V = M,\label{Eq00}
\end{equation}
where $M$ is an integration constant.

The variation of the action with respect to the metric $ g^{\mu
\nu}$ gives

\begin{equation}
 R_{\mu\nu} (\Gamma) ( \frac{-\Phi}{\kappa} + 2 \epsilon R  \sqrt{-g}) +
\Phi \frac{1}{2} \phi,_{\mu} \phi,_{\nu} - \frac{1}{2}(\epsilon
R^{2} + U(\phi)) \sqrt{-g} g_{\mu\nu} = 0. \label{Eq01}
\end{equation}

It is interesting to notice that if we contract this equation with
 $ g^{\mu \nu}$ , the $\epsilon$ terms do not contribute. Solving the
 scalar curvature from this and inserting in the other
$\epsilon$ - independent equation $L_{1} = M$  we get the solution
for the ratio of the measures i.e. $\chi = \frac{\Phi}{\sqrt{-g}}
= \frac{2U(\phi)}{M+V(\phi)}$.


We see that under a conformal transformation given by

\begin{equation}
\overline{g}_{\mu\nu}   = (\frac{\omega}{\sqrt{-g}}) g_{\mu\nu}
 = (\chi -2\kappa \epsilon R) g_{\mu\nu},
\end{equation}
the new metric  $\overline{g}_{\mu\nu} $ defines the known
Einstein frame. Equations (\ref{Eq01}) can now be expressed in
this frame as

\begin{equation}
\overline{R}_{\mu\nu} -  \frac{1}{2}\overline{g}_{\mu\ \nu}
\overline{R} = \frac{\kappa}{2} T^{eff}_{\mu\nu}, \label{Eq02}
\end{equation}
where
\begin{equation}
 T^{eff}_{\mu\nu} =
\frac{\chi}{\chi -2 \kappa \epsilon R} (\phi_{,\mu} \phi_{,\nu} -
\frac{1}{2} \overline {g}_{\mu\nu} \phi_{,\alpha} \phi_{,\beta}
\overline{g}^{\alpha\beta}) + \overline{g}_{\mu\nu} V_{eff},
\label{Eq03}
\end{equation}
with
\begin{equation}
 V_{eff}  = \frac{\epsilon R^{2} + U}{(\chi -2 \kappa \epsilon R)^{2}}+ \lambda. \label{Eq04}
\end{equation}
The constant $\lambda$ that appears in this latter expression
could be obtained from the original lagrangian, where in it is
added a term proportional to $\lambda \Omega^4$, where $\Omega$ is
the conformal factor. This renders the action to be
\begin{equation}
S_{eff,\Lambda
}=\int\sqrt{-\overline{g}}d^{4}x\left[-\frac{1}{\kappa}\overline{R}(\overline{g})
+p\left(\phi,R\right) \right], \label{act.lambda}
\end{equation}
where
\begin{equation}
 p(\phi,R) = \frac{\chi}{\chi -2 \kappa \epsilon R}\frac{1}{2}\overline{g}^{\,\alpha\beta}\phi_{,\alpha}\phi_{,\beta}
  - V_{eff},
\end{equation}
with $V_{eff}$ given by Eq. (\ref{Eq04}).

Also, Eq.(\ref{Eq00}) expressed in terms of $
\overline{g}^{\alpha\beta}$ becomes $\frac{-1}{\kappa} R(\Gamma,g)
+ (\chi -2\kappa \epsilon R) \frac{1}{2}
\overline{g}^{\mu\nu}\partial_{\mu} \phi \partial_{\nu} \phi - V =
M$, where we have written $V$ in place of $V(\phi)$. This allows
us to solve for $R$ and thus we get
\begin{equation}
R = \frac{-\kappa (V+M) +\frac{\kappa}{2}
\overline{g}^{\mu\nu}\partial_{\mu} \phi \partial_{\nu} \phi \chi}
{1 + \kappa ^{2} \epsilon \overline{g}^{\mu\nu}\partial_{\mu} \phi
\partial_{\nu} \phi}. \label{Eq05}
\end{equation}

Note that when we insert Eq. (\ref{Eq05}) into Eq. (\ref{Eq04}),
it depends on the derivatives of the scalar field. It acts as a
normal scalar field potential under the conditions of slow rolling
or low gradients and in the case the scalar field is near the
region $M+V = 0$.

In the scale invariant case, where $V$ and $U$ are given by
$V(\phi) = f_{1}  e^{\alpha\phi}$ and $ U(\phi) =  f_{2}
e^{2\alpha\phi}$, respectively\cite{TMT2}, it is interesting to
study the shape of $ V_{eff} $ as a function of $\phi$ in the case
when $\phi$ change a little bit. In this case, $ V_{eff}$ can be
regarded as a scalar field potential. Then from Eq. (\ref{Eq05})
we get $R = -\kappa (V+M)$, which inserted into Eq. (\ref{Eq04})
gives,
\begin{equation}\label{effpotslow}
 V_{eff}  =
\frac{(f_{1} e^{ \alpha \phi }  +  M )^2}{4(\epsilon \kappa
^2(f_{1}e^{\alpha \phi}  +  M )^{2} + f_{2}e^{2 \alpha \phi })}+
\lambda.
\end{equation}

\begin{figure}[th]
\includegraphics[width=5.0in,angle=0,clip=true]{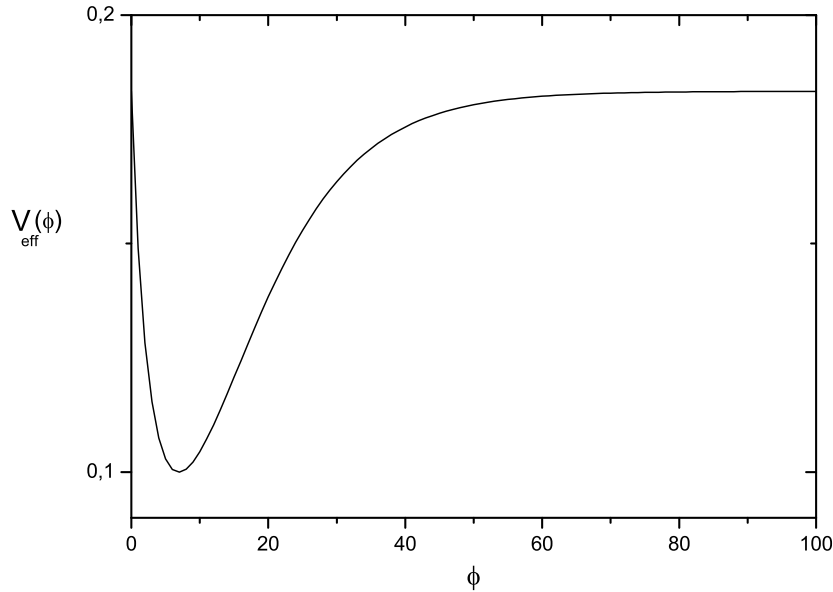}
\caption{ The form of effective potential $V_{eff}(\phi)$ versus
the scalar field $\phi$. We have used  $M=-1$, $\epsilon=-1$, $f_1
= 1/2$, $f_2=1$, $\lambda=1/10$ and $\kappa=1$.}\label{fig1}
\end{figure}

In Fig.\ref{fig1} we have plotted the effective potential,
$V_{eff}$ as a function of the scalar field, $\phi$ for $M=-1$,
$\epsilon=-1$, $f_1 = 1/2$, $f_2=1$, $\lambda=1/10$ and
$\kappa=1$. We should mention here that the choices of the values
of the different parameters was done in virtue that they respect
the conditions for a stable emerging universe. Notice that it
shows a flat region of positive vacuum energy for large $\phi$, a
minimum obtained at zero without fine tuning.

In the limit $ \alpha\phi \rightarrow \infty $ the effective
scalar potential, $ V_{eff} $, becomes $V_{eff} \rightarrow
\frac{f_{1}^{2}}{4(\epsilon \kappa ^{2} f_{1}^{2} +
f_{2})}+\lambda $.

Notice that in all the above discussion it is fundamental that $
M\neq 0$. If $M = 0$ the potential becomes just a flat one,
$V_{eff} = \frac{f_{1}^{2}}{4(\epsilon \kappa ^{2} f_{1}^{2} +
f_{2})}+ \lambda$ everywhere (not only at high values  of $\alpha
\phi$). All the non trivial features necessary for a graceful
exit, the other flat region associated to the Planck scale and the
minimum at zero if $M<0$ are all lost. The case $ M\neq 0$ implies
that we are considering a situation with S.S.B. of scale
invariance.

We now want to consider the detailed analysis of emerging universe
solutions. We start considering the closed
Friedmann-Robertson-Walker metric,
\begin{equation}
ds^2 =dt^2 - a(t)^2 \left(\frac{dr^2}{1 -r^2}+ r^2(d\theta^2
+sin^2\theta d\phi^2)\right),
\end{equation}
where $a(t)$ is the scale factor, $t$ represents the cosmic time
and due to homogeneity and isotropy we take the scalar field
$\phi$ to be a function of the cosmological time only, i.e.
$\phi=\phi(t)$.

We will consider a scenario where the scalar field $\phi$ is
moving in the extreme right region, i.e. for asymptotic values of
$\phi \,(\phi \rightarrow \infty) $. In this case, the expressions
for the energy density $\rho$ and pressure $p$ are given by,
\begin{equation}\label{Eq06}
\rho = \frac{A}{2}(1+\kappa^2 \epsilon \dot{\phi}^2) \dot{\phi}^2
+ (C+B\dot{\phi}^4),
\end{equation}
and
\begin{equation}
p = \frac{A}{2}(1+\kappa^2 \epsilon \dot{\phi}^2) \dot{\phi}^2 -
(C+B\dot{\phi}^4), \label{Eq07}
\end{equation}
where the constants $A,B$ and $C$ are given by
\begin{eqnarray}\label{A}
A &=& \frac{f_2}{f_2 + \kappa^2\epsilon f_1^2}\,,\\
\end{eqnarray}
\begin{eqnarray}
B &=& \frac{\epsilon\kappa^2}{4(1+\kappa^2\epsilon f_1^2/f_2)}
= \frac{\epsilon \kappa^2}{4}\,A \,,\label{B} \\
\end{eqnarray}
and
\begin{eqnarray}
C &=& \frac{f_1^2}{4\,f_2(1+\kappa^2\epsilon f_1^2/f_2)} =
\frac{f_1^2}{4f_2}\,A\,\label{C}-\frac{1}{4 \epsilon
\kappa^2}+\lambda,
\end{eqnarray}
respectively.

It is interesting to notice that all terms proportional to
$\dot{\phi}^4$ behave like "radiation", since
\begin{equation}
p_{\dot{\phi}^4} = \frac{A}{2}\kappa^2 \epsilon \dot{\phi}^4 -
B\dot{\phi}^4 = (\frac{A}{2}\epsilon \kappa^2 -B)\dot{\phi}^4 =
\frac{A}{4}\kappa^2 \epsilon \dot{\phi}^4,
\end{equation}
and
\begin{equation}
\rho_{\dot{\phi}^4} = \frac{A}{2}\kappa^2 \epsilon \dot{\phi}^4 +
B\dot{\phi}^4 = (\frac{A}{2}\epsilon \kappa^2 + B)\dot{\phi}^4 =
\frac{3A}{4}\kappa^2 \epsilon \dot{\phi}^4,
\end{equation}
so that $p_{\dot{\phi}^4} = \frac{\rho_{\dot{\phi}^4} }{3}$ is satisfied.
This is also a consistency check, since the $\epsilon$ terms are associated
to a conformally invariant $R^2$ term (in the first order formalism) and in
the region $\phi \rightarrow \infty$ that symmetry should hold, at least for
those contributions. The $\epsilon$ terms therefore do not contribute to the
trace of the energy momentum tensor in the region $\phi \rightarrow \infty$.

The equations that determines such static universe $a(t) = a_0
=constant$, $\dot{a}=0$ and $\ddot{a}=0$ give rise to a
restriction on $\dot{\phi}_0$. Since $\ddot{a}=0$ and it is
proportional to $\rho + 3p$, we must require that $\rho + 3p = 0$,
which leads to

\begin{equation}\label{e1}
(B - A\kappa^2 \epsilon)\dot{\phi}^4_0 - A\dot{\phi}^2_0 +C=0.
\end{equation}
This equation has two roots, the first being given by
\begin{equation}\label{e2}
\dot{\phi}_0^2 = \dot{\phi}_1^2=\frac{-2\sqrt{f_2} -
\sqrt{\Delta}}{3\sqrt{f_2}\epsilon}\,,
\end{equation}
 and the second one is
\begin{equation}
\dot{\phi}_0^2 = \dot{\phi}_2^2=\frac{-2\sqrt{f_2} +
\sqrt{\Delta}}{3\sqrt{f_2}\epsilon}\,\label{eq09},
\end{equation}
where
$$
\Delta = f_2 + 12f_2\lambda \epsilon + 12 \lambda f_1^2
\epsilon^2.
$$
From these expression we could get some specific range of the
parameters in order that either $\rho_0$ and $\dot{\phi_0}^2$
become positive. In the following section we will restrict these
parameters in order to describe an emerging universe.

\section{Stability of the static solution in the scale invariant $R^2$ model}

We will now consider the perturbation equations. Considering small
deviations of $\dot{\phi}$  from the static emerging solution
value $\dot{\phi}_0$ and also considering the perturbations of the
scale factor $a$, we obtain, from Eq.~(\ref{Eq06}), that

\begin{equation}\label{eq.density-pert.}
\delta \rho = A \dot{\phi}_0 \delta \dot{\phi} +4( B +
\frac{1}{2}\kappa^2 \epsilon A )\dot{\phi}_0^3 \delta \dot{\phi}.
\end{equation}
At the same time $\delta \rho$ can be obtained from the
perturbation of the Friedmann equation $
3(\frac{1}{a^2}+H^2)=\kappa \rho$, which gives under a
perturbation respect to the background static solution
\begin{equation}\label{pert.Fried.eq.}
-\frac{6}{a_0^3}\delta a =\kappa \delta \rho.
\end{equation}

On the other hand, from the second order Friedmann equation, $
\frac{1+\dot{a}^2 + 2a\ddot{a}}{a^2}=-\kappa p $, we get that
\begin{equation}
\frac{2}{a_0^2} = -2\kappa p_0 = \frac{2}{3}\kappa \rho_0=
\Omega_0 \kappa \rho_0,
\end{equation}
where we have used $p_0=-\rho_0/3$ at $a=a_0$,  and we have chosen
to express our result in terms of $\Omega_0$, defined by
$p_0=(\Omega_0-1)\rho_0$, which for the emerging solution has the
value $\Omega_0=\frac{2}{3}$. Using this in Eq.
(\ref{pert.Fried.eq.}), we obtain
\begin{equation}\label{pert.Fried.eq.2}
\delta \rho = -\frac{3\Omega_0 \rho_0}{a_0}\delta a.
\end{equation}
Substituting this latter expression into Eq.
(\ref{eq.density-pert.}) yields to a linear relation between
$\delta \dot{\phi}$ and $\delta a$ so that
\begin{equation}\label{delta-delta}
\delta \dot{\phi}=D_0\delta a,
\end{equation}
where
\begin{equation}
D_0 = -\frac{3\Omega_0 \rho_0}{a_0 \dot{\phi}_0 (A + 4( B +
\frac{1}{2}A \kappa^2 \epsilon \dot{\phi}_0^2))}.
\end{equation}

Since we could write that $p=(\Omega-1)\rho$, with
\begin{equation}\label{Omega-eq.}
\Omega = 2\Big(1 - \frac{V_{eff}}{\rho}\Big),
\end{equation}
where,
\begin{equation}\label{V-eq.}
V_{eff} =C + B\,\dot{\phi}^4,
\end{equation}
and therefore,  the perturbation of the second order Friedmann
equation leads to,
\begin{equation}\label{pert.Fried.eq.2}
-\frac{2\delta a}{a_0^3}+2\frac{\delta\ddot{a}}{a_0}=-\kappa
\delta p =-\kappa \delta ((\Omega-1)\rho),
\end{equation}
to evaluate this, we use (\ref{Omega-eq.}), (\ref{V-eq.}) and the
expressions that relate the variations in $a$ and $\dot{\phi}$
(\ref{delta-delta}).  Defining the "small"  variable $\beta$ as
$a(t) = a_0( 1+ \beta) $ we obtain, $2\ddot{\beta}(t) +
W_0^2\beta(t) = 0\,, $ where,
\begin{equation}
W_0^2 = \frac{4 \kappa}{3}\left[ \frac{12\,B\,\dot{\phi}_0^2}{A +
\dot{\phi}_0^2\,(4B + 2\kappa^2\epsilon A)}  - \frac{3}{\rho_0}(C
+ B \dot{\phi}_0^4)\right],\label{Eq08}
\end{equation}
with $ \rho_0 = A(1 +
\kappa^2\,\epsilon\,\dot{\phi}_0^2)\frac{\dot{\phi}_0^2}{2} + C +
B\,\dot{\phi}_0^4 $.

Stability of the static solution requires that  $W_0^2
>0$. This static solution has to respect the
conditions $\dot{\phi}_0^2>0$ and $\rho_0>0$. Only the second
solution, i.e. $\dot{\phi}_2^2$ (see Eq. (\ref{eq09})), fulfills
these requirements, providing that the parameters satisfy the
following inequalities

\begin{eqnarray}
\epsilon &<& 0 \label{cc1}\\
-\epsilon f_1^2  & < & f_2\label{cc2} \\
0< &\lambda& < - \frac{1}{12\epsilon\left(1+\frac{\epsilon f_1^2
}{f_2}\right)},\label{cc3}
\end{eqnarray}
Here, we have taken $\kappa = 1$.

\section{Stability of the Static Solution: Dynamical System.}

 The study of the stability of the static solution and the
properties of the different equilibrium points could be done in a
more systematic way by using a dynamical system approach.
In this scheme we rewrite the Friedmann  and the conservation of
energy equations as an autonomous system in terms of the variables
$H$ and $x \equiv \dot{\phi}^2$.
In order to do so, we differentiating the Fridmann equation and
after using the expressions for $\rho$ and $p$, Eq.~(16, 17), and
the energy conservation equation we obtain:

\begin{equation}
\dot{H} = \frac{1}{a^2} - \frac{\kappa}{2}\,A\,(1 +
\kappa^2\,\epsilon \,x)x \,.\label{dinamic0}
\end{equation}

We can rewrite Eq.~(\ref{dinamic0}) and the energy conservation
equation as the following two equations.

\begin{eqnarray}
\dot{H} &=& \frac{\kappa}{3}\Big[C + B\, x^2 - A(1 +
\kappa^2\,\epsilon\,x)\,x\Big] - H^2, \label{dinamic1}\\
\nonumber \\
\dot{x} &=& - \frac{3A\,(1 + \kappa^2\,\epsilon
\,x)\,x}{\frac{A}{2} + A\,\kappa^2\,\epsilon \,x + 2B\,x}\,H \,\,,
\label{dinamic2}
\end{eqnarray}
where we have eliminated $a$ from Eq.~(\ref{dinamic0}) by using
the Friedmann equation and we have used the definition of the
variable $x$ in order to get Eq.~(\ref{dinamic2}) from the energy
conservation equation.
The equations (\ref{dinamic1}) and (\ref{dinamic2}) are a
two-dimensional autonomous system on the variables $H$ and $x$.

In order to study the stability of the static solutions we look
for critical points of the system (\ref{dinamic1}) and
(\ref{dinamic2}). These points are

\begin{eqnarray}
\Big\{ H = 0 , &x& = \frac{A - \sqrt{A^2 + 4(A\epsilon - B)C}}{2(B
- A\epsilon)}\Big\} ; \label{crit1}\\  \Big\{ H = 0 , &x& =
\frac{A + \sqrt{A^2
+ 4(A\epsilon - B)C}}{2(B - A\epsilon)}\Big\}\,; \label{crit2}\\
\Big\{ H = - \sqrt{\frac{C}{3}} \,, \,\,x = 0 \Big\} ; \Big\{ H =
\sqrt{\frac{C}{3}} \,, \,\,x = 0 \Big\}\,&;& \Big\{ H = -
\sqrt{\frac{\frac{B}{\epsilon^2} + C}{3}} \,, \,\,x = -
\frac{1}{\epsilon}\Big\} ;\\
 \Big\{ H =
\sqrt{\frac{\frac{B}{\epsilon^2} + C}{3}} \,, \,\,x = -
\frac{1}{\epsilon}\Big\}\,. \label{crit3}
\end{eqnarray}

The critical points have different properties depending on the
values of the parameters of the model ($f_1, f_2, \epsilon$,
$\lambda$). At this moment we are not going to give an exhaustive
description of these properties for all the critical points,
instead, we are going to focus on the particular critical points
which are related with static universe.
From the definition of the variables $H$ and $x$ we can note that
only the first two critical points Eqs.~(\ref{crit1}, \ref{crit2})
correspond to a static universe.
In order to study the nature of these two critical points we
linearize the equations (\ref{dinamic1}) and (\ref{dinamic2}) near
these critical points. From the study of the eigenvalues of the
system we found that the first critical  point, Eq.~(\ref{crit1}),
could be a center or a saddle point, depending on the values of
the parameters of the model.
On the other hand the second critical point Eq.~(\ref{crit2}) is a
saddle.

Stable static solutions correspond to a center. Then, by impose
the critical point, Eq.(\ref{crit1}), becomes a center, we recover
the stability condition Eqs.(\ref{cc1}), (\ref{cc2}) and
(\ref{cc3}).

%
%
In order to satisfy the requirements of stability we are going
take the values $\epsilon = -1$, $f_1 = 1/2$, $f_2 = 1$ and
$\lambda = 1/10$.

\begin{figure}[h]
\begin{center}
\includegraphics[width=2.5in,angle=0,clip=true]{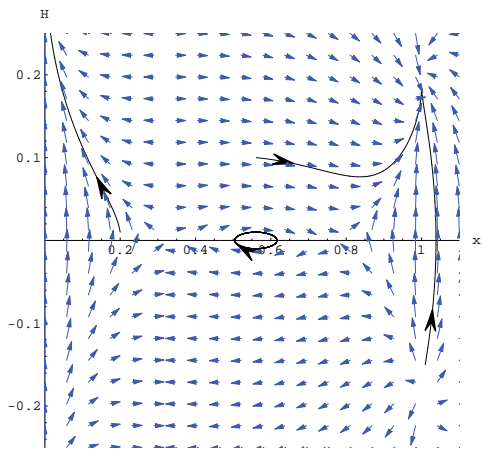}
\caption{Plot showing part of the \textit{Direction Field} of the
system and four numerical  solutions.} \label{fase1}
\end{center}
\end{figure}

 In Fig.~\ref{fase1} it is shown a phase portrait for four
numerical  solution to Eqs.~(\ref{dinamic1}) and (\ref{dinamic2}).
Also, in this figure we have included the \textit{Direction Field}
of the system in order to have a picture of what a general
solution look like.
In this figure are four of the six critical points described in
Eqs.~(\ref{crit1})-(\ref{crit3}). One of this point is the center
equilibrium point ($H = 0, x = 0.56$), the saddle point ($H = 0, x
= 0.77$), the point ($H = 0.18, x = 1)$ is a future attractor and
($H = - 0.18, x = 1)$ is a past attractor. The other equilibrium
points are $(H = 0.38, x = 0)$ and $(H = -0.38, x = 0)$ which are
a future attractor point and a past attractor point respectively.

\begin{figure}[h]
\begin{center}
\includegraphics[width=2.5in,angle=0,clip=true]{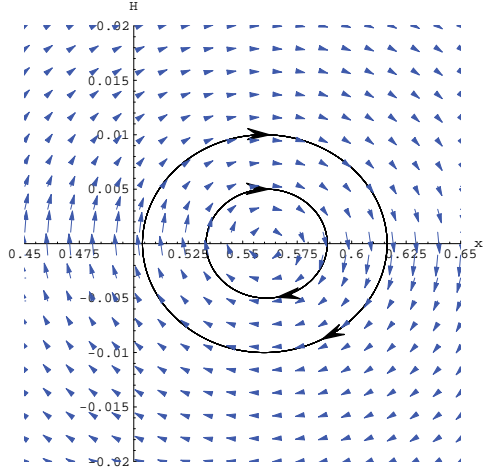}
\caption{Plot showing the \textit{Direction Field} near the center
critical point and two numerical solution.} \label{fase2}
\end{center}
\end{figure}

As we have mentioned, only critical points with $H = 0$ represent
a static universe. Then in our case only the critical point which
is a center correspond to a solution which represent a static and
stable universe.
In Fig.~\ref{fase2} we show the \textit{Direction Field} near this
critical point ($H = 0, x = 0.56$) together with two numerical
solution.

\section{Inflation and its Graceful Exit}
The emerging phase owes its existence to a strictly constant
vacuum energy (which here is represented by the value of $A$) at
very large values of the field $\phi$. In fact, while for $M=0$
the effective potential of the scalar field is perfectly flat, for
any $M \neq 0$ the effective potential acquires a non trivial
shape.

We consider then the relevant equations for the model in the slow
roll regime, i.e. for $\dot{\phi}$ small and when the scalar field
$\phi$ is large, but finite and we consider the first corrections
to the flatness to the effective potential. Dropping higher powers
of $\dot{\phi}$ in the contributions for the kinetic energy and in
the scalar curvature $R$, we obtain
\begin{eqnarray}
\rho =  \frac{1}{2}\gamma\dot{\phi}^2 + V_{eff}, \label{energydensityslow}\\
\gamma = \frac{\chi}{\chi -2\kappa\epsilon R}, \label{gamma}\\
R = -\kappa (V+M)\label{R}.
\end{eqnarray}
Here, as usual  $\chi= \frac{2U(\phi)}{M+V(\phi)}$. In the slow
roll approximation, we can drop the second derivative term of
$\phi$ and the second power of $\dot{\phi}$ in the equation for
$H^2$ and we get
\begin{eqnarray}
3H \gamma \dot{\phi} = - V^{\prime}_{eff} \label{slowroll},\\
3H^2 = \kappa V_{eff} \label{Friedmannslow},
\end{eqnarray}
where $ V^{\prime}_{eff}= \frac{dV_{eff}}{d\phi}$. The relevant
expression for $V_{eff}$ will be that given by (\ref{effpotslow}),
i.e., where all higher derivatives are ignored in the potential,
consistent with the slow roll approximation.

We now display the relevant expressions for the region of very
large, but not infinite $\phi$, these are:
\begin{eqnarray}
 V_{eff} = C + C_1 exp(-\alpha \phi),\label{effpotatlargephi}\\
\chi =2 \frac{f_2}{f_1}exp(\alpha \phi)-
2M\frac{f_2}{f^2_1},\label{chiatlargeph}
\end{eqnarray}
and
\begin{eqnarray} \gamma = \gamma_0 + \gamma_1 exp(-\alpha
\phi). \label{gammaatlargephi}
\end{eqnarray}

The relevant constants that will affect our results are, $C$, as
given by (\ref{C}) and $C_1 $ and $ \gamma_0$ given by
\begin{eqnarray}
  C_1 = -\frac{8\epsilon \kappa^2 f^3_1 M}{(4f_2+4\kappa^2\epsilon f_1^2)^2}+
  \frac{2 f_1 M}{4f_2+4\kappa^2\epsilon f_1^2},\label{C1}\end{eqnarray}
and
  \begin{eqnarray}
 \gamma_0 = \frac{ f_2}{f_2+\kappa^2\epsilon f_1^2}, \label{gamma0}
\end{eqnarray}
respectively.

Using Eq. (\ref{effpotatlargephi}) we can calculates the key
landmarks of the inflationary history: first, the value of the
scalar field where inflation ends, $\phi_{end}$ and a value for
the scalar field $\phi_{*}$ bigger than this ($\phi_{*} >
\phi_{end}$) and which happens earlier, which represents the
"horizon crossing point". We must demand then that a typical
number of e-foldings, like $N=60$, takes place between $\phi_{*}$,
until the end of inflation at $\phi=\phi_{end}$.

To determine the end of inflation, we consider the quantity
$\delta = - \frac{\dot{H}}{H^2}$ and consider the point in the
evolution of the Universe where $\delta =1$, only when $\delta <
1$, we have an accelerating Universe, so the point $\delta =1$
represents indeed the end of inflation. Calculating the derivative
with respect to cosmic time of the Hubble expansion using
(\ref{Friedmannslow}) and (\ref{slowroll}), we obtain that the
condition $\delta =1$ gives

\begin{equation}
 \delta =\frac{1}{2\gamma}(V^{\prime}_{eff}/V_{eff})^2 = 1, \label{endofinf.eq.}
\end{equation}
 working to leading order, setting $\gamma = \gamma_0$, $V_{eff} = C $
 and $V^{\prime}_{eff}= - \alpha C_1 exp(-\alpha \phi_{end})$, this gives as a solution,
\begin{equation}
 exp(\alpha \phi_{end})= -\frac{\alpha C_1 }{C \sqrt{2\kappa
 \gamma_0}},
 \label{sol.endofinf.eq.}
\end{equation}
notice that if $M$ and $f_1$ have different signs and if
$\epsilon<0$, $C_1<0$ for the allowed range of  parameters the
stable emerging solution, so $-C_1$ represents the absolute value
of $C_1$. We now consider $\phi_{*}$ and the requirement that this
precedes $\phi_{end}$ by $N$ e-foldings,
\begin{equation}
N= \int Hdt = \int \frac{H}{\dot{\phi}} d\phi = -\int
\frac{3H^2\gamma}{ V^{\prime}_{eff}} d\phi, \label{N}
\end{equation}
where in the last step we have used the slow roll equation of
motion for the scalar field (\ref{slowroll}) to solve for
$\dot{\phi}$. Solving $H^2$ in terms of $V_{eff}$ using
(\ref{Friedmannslow}), working to leading order, setting $\gamma =
\gamma_0$ and integrating, we obtain the relation between
$\phi_{*}$ and  $\phi_{end}$,
\begin{equation}\label{sol.crossing.}
exp(\alpha \phi_{*}) = exp(\alpha \phi_{end}) - \frac{N \alpha^2
C_1 }{C\kappa \gamma_0},
\end{equation}
as we mentioned before $C_1<0$ for the allowed range of
parameters the stable emerging solution, so that $\phi_{*} >
\phi_{end}$ as it should be for everything to make sense.
Introducing Eq. (\ref{sol.endofinf.eq.}) into Eq.
(\ref{sol.crossing.}), we obtain,
\begin{equation}\label{sol.crossing.final}
exp(\alpha \phi_{*}) = -\frac{ C_1 }{C\sqrt{\kappa}}(
\frac{\alpha} {\sqrt{2\gamma_0}} +
\frac{N\alpha^2}{\sqrt{\kappa}\gamma_0} ).
\end{equation}

We finally calculate the power of the primordial scalar
perturbations. If the scalar field $\phi$ had a canonically
normalized kinetic term, the spectrum of the primordial
perturbations will be given by the equation
 \begin{equation}\label{primordialpert.}
 \frac{\delta \rho }{\rho} \propto \frac{H^2}{\dot{\phi}},
 \end{equation}
however, as we can see from (\ref{energydensityslow}), the kinetic
term is not canonically normalized because of the factor $\gamma$
in that equation.

In this point we will study the scalar and tensor perturbations
for our model where the kinetic term is not canonically
normalized. The  general expression for the perturbed metric about
the  Friedmann-Robertson-Walker is
\begin{eqnarray*}
ds^2&=&-(1+2 F) dt^2+ 2 a(t) D_{,\,i}dx^i dt + a^2(t) [(1-2 \psi)
\delta_{ij}+2 E_{,i,j}+2 h_{ij}] dx^i dx^j,
\end{eqnarray*}
where  $F$, $D$, $\psi$ and $E$ are the scalar type metric
perturbations and $h_{ij}$ characterizes the transverse-traceless
tensor perturbation. The power spectrum of the curvature
perturbation in the slow-roll approximation for a not-canonically
kinetic term becomes Ref.\cite{Garriga}(see also Refs.\cite{p1})

\begin{equation}
P_S = k_1\left(\frac{\delta \rho
}{\rho}\right)^2=k_1\,\frac{H^2}{c_s\,\delta},\label{pec}
\end{equation}
where it was  defined  "speed of sound", $c_s$, as
$$
c_s^2=\frac{P_{,\,X}}{P_{,\,X}+2XP_{,\,XX}},
$$
 with $P(X,\phi)$  an  function of the scalar field and of the
 kinetic term $X=-(1/2)g^{\mu\nu}\partial_\mu \phi\partial_\nu \phi$.
 Here $P_{,\,X}$ denote the derivative with respect $X$.
 In our case $P(X,\phi)=\gamma(\phi)\,X-V_{eff}$, with $X=\dot{\phi}^2/2$.
 Thus, from eq.(\ref{pec}) we get
\begin{equation}
P_S=k_1\frac{H^4}{\gamma(\phi) \dot{\phi}^2}.\label{eq10}
\end{equation}

 The scalar spectral index $n_s$, is defined by
\begin{equation}
n_s-1=\frac{d\ln P_S}{d\ln k}=-2\delta-\eta-s,\label{ns}
\end{equation}
where $\eta=\frac{\dot{\delta}}{\delta\,H}$ and
$s=\frac{\dot{c_s}}{c_s\,H}$, respectively.

On the other hand, the generation of tensor perturbations during
inflation  would produce gravitational wave. The amplitude of
tensor perturbations was evaluated in Ref.\cite{Garriga}, where
$$
P_T=\frac{2}{3\pi^2}\,\left(\frac{2XP_{,\,X}-P}{M_{Planck}^4}\right),
$$
and the tensor spectral index $n_T$, becomes
$$
n_T=\frac{d\ln P_T}{d\ln k}=-2\delta,
$$
and they satisfy a generalized consistency relation
\begin{equation}
r=\frac{P_T}{P_S}=-8\,c_s\,n_T. \label{ration}
\end{equation}

\begin{figure}[th]
\includegraphics[width=5.0in,angle=0,clip=true]{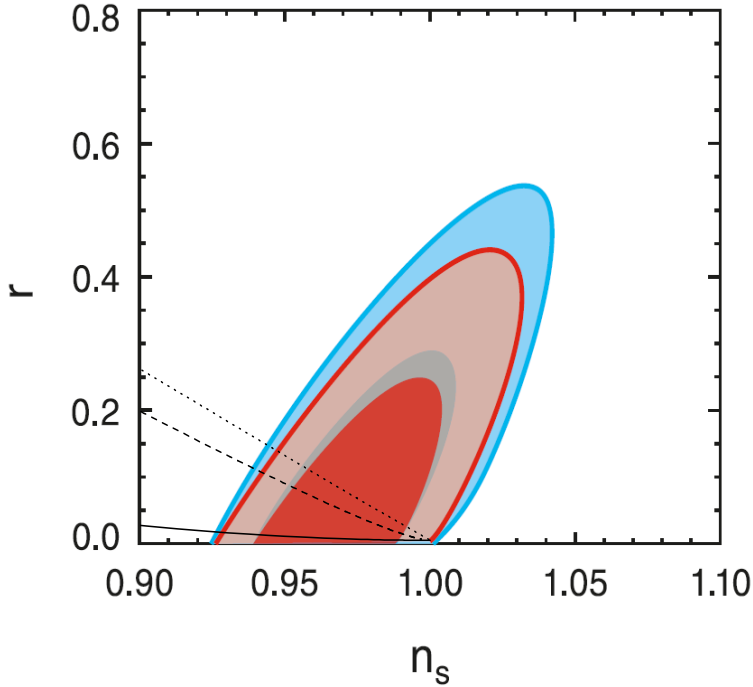}
\caption{ The plot shows $r$ versus $n_s$ for three values of
$\alpha$. For $\alpha=1$ solid line,  $\alpha=0.1$ dash line and
$\alpha=0.01$ dots line, respectively. Here, we have fixed the
values $M=-1$, $\epsilon=-1$, $f_1 = 1/2$, $f_2=1$, $\lambda=1/10$
and $\kappa=1$, respectively. The seven-year WMAP data places
stronger limits on the tensor-scalar ratio (shown in red) than
five-year data (blue) \cite{Larson:2010gs}.\label{fig2}}
\end{figure}

Therefore, the scalar field (to leading order) that should appear
in Eq. (\ref{primordialpert.}) should be $\sqrt{\gamma_0}\phi$ and
instead of Eq. (\ref{eq10}) , we must use
\begin{equation}\label{primordialpert.corr}
 \frac{\delta \rho }{\rho} =
 \frac{H^2}{\sqrt{\gamma_0}\dot{\phi}}.
 \end{equation}
The power spectrum of the perturbations goes, up to a factor of
order one, which we will denote $k_1$ as $(\delta \rho /\rho )^2$,
so we have,
\begin{equation}\label{primordialpert.power}
P_S = k_1\left(\frac{\delta \rho }{\rho}\right)^2 =
k_1\frac{H^4}{\gamma_0 \dot{\phi}^2},
 \end{equation}
this quantity should be evaluated at $\phi=\phi_{*}$ given by
(\ref{sol.crossing.final}). Solving for $\dot{\phi}$ from the slow
roll equation (\ref{slowroll}), evaluating the derivative of the
effective potential using (\ref{effpotatlargephi}) and solving for
$H$ from (\ref{Friedmannslow}), we obtain, to leading order,
\begin{equation}\label{primordialpert.power-almostfinal}
P_S = k_1\frac{\kappa^3 \gamma_0 C^3 }{3\alpha^2 C_1}
exp(2\alpha\phi_{*}),
\end{equation}
using then (\ref{sol.crossing.final}) for $\exp(\alpha\phi_{*})$,
we obtain our final result,
\begin{equation}\label{primordialpert.power-final}
P_S = k_1\frac{\kappa^2 C }{3}(\frac{1}{\sqrt{2}} + \frac{N
\alpha}{\sqrt{\gamma_0 \kappa}})^2,
\end{equation}
it is very interesting first of all that $C_1$ dependence has
dropped out and with it all dependence on $M$. In fact this can be
regarded as a non trivial consistency check of our estimates,
since apart from its sign, the value $M$ should not affect the
results. This is due to the fact that from a different value of
$M$  (although with the same sign), we can recover the original
potential by performing a shift of the scalar field $\phi$.

In Fig.\ref{fig2} we show the dependence of the tensor-scalar
ratio $r$ on the spectral index $n_s$. From left to right
$\alpha=1$ (solid line), $\alpha=0.1$ (dash line) and
$\alpha=0.01$ (dots line), respectively. From
Ref.\cite{Larson:2010gs}, two-dimensional marginalized
 constraints (68$\%$ and 95$\%$ confidence levels) on inflationary parameters
$r$ and $n_s$, the spectral index of fluctuations, defined at
$k_0$ = 0.002 Mpc$^{-1}$. The seven-year WMAP data places stronger
limits on $r$ (shown in red) than five-year data
(blue)\cite{WMAP5}. In order to write down values that relate
$n_s$ and $r$, we used Eqs.(\ref{ns}) and (\ref{ration}).  Also we
have used  the values $M=-1$, $\epsilon=-1$, $f_1 = 1/2$, $f_2=1$,
$\lambda=1/10$ and $\kappa=1$, respectively.

From Eqs.(\ref{N}), (\ref{ns}) and (\ref{ration}), we observed
numerically that for $\alpha = 1$, the curve $r = r(n_s)$ (see
Fig.\ref{fig2}) for WMAP 7-years enters the 95$\%$ confidence
region where the ratio $r\simeq 0.011$, which corresponds to the
number of e-folds, $N \simeq 32$. For $\alpha=0.1$, $r \simeq
0.103$ corresponds to $N \simeq 227$ and for $\alpha=0.01$, $r
\simeq 0.136$ corresponds to $N \simeq 14137$. From 68$\%$
confidence region for $\alpha = 1$, $r\simeq 0.010$, which
corresponds to $N\simeq 34$. For $\alpha = 0.1$, $r \simeq 0.08$
corresponds to $N \simeq 240$ and for $\alpha=0.01$, $r \simeq
0.109$ corresponds to $N \simeq 14279 $.  We noted that the
parameter $\alpha$, which lies in the range $1 > \alpha > 0$, the
model is well supported by the data as could be seen from
Fig.\ref{fig2}.

\section{Discussion and Conclusions}
The consideration of a scale invariant action leads to exactly the
type of scalar field potentials needed to obtain an emerging
universe scenario. The addition of a $R^{2}$-term gives rise to
solutions where $\epsilon <0$, $-\epsilon f_1^2 < f_2$ and $0 <
\lambda <  - \frac{1}{12\epsilon\left(1+\frac{\epsilon f_1^2
}{f_2}\right)}$ and to the stability of the initial Einstein
universe solution.

We find that this is accomplished because the $\epsilon R^{2}$
term is shown to have the effect of introducing nonlinearities in
the theory that stabilize the solutions.

In addition  in order for the whole idea to work we need to take
$M \neq 0$, which means we are considering the S.S.B. of the
theory and this is what gives the effective potential a non
trivial shape, which allows a graceful exit from the inflationary
phase and can give the adequate power of the density
perturbations.

It should be pointed out also that the $R^{2}$ theory studied
here, in the context of an action that contains a measure $\Phi$
independent of the metric and with the use of the first order
formalism (the connections are considered as independent degrees
of freedom in the action principle), leads to equations of motion
that are only second order, i.e. only second derivatives of the
metric and the matter fields appear (although higher powers of the
derivatives of the scalar field do appear). This after the new
measure $\Phi$ and the connections are solved (through the
equations of motion) in terms of the metric and matter fields and
after we express our results in the Einstein frame.

This is in contrast to the usual $R^{2}$ theories in the second
order formalism and with standard measure everywhere in the
action. In this case, these theories lead to fourth order
equations for the metric field. In Ref.\cite{barrow} it was shown
that the fourth order structure of the equations can be
reformulated as a system of second order equations which contain
however one additional degree of freedom, a scalar field. For
studies of inflation in the usual $R^{2}$ theories in the second
order formalism, see Ref.\cite{mijic}. In the case of the  usual
$R^{2}$ theories this scalar field contains a potential with a
flat region, as shown in Fig.\ref{fig1}.

Using the WMAP seven-year data, we have found some constraints for
the parameters appearing in our model. In particular,
Fig.\ref{fig2} shows that for the values of the parameter $1>
\alpha > 0$, the model is well supported by the WMAP data.

\section{Acknowledgements}
This work was supported by Comision Nacional de Ciencias y
Tecnolog\'{\i}a through FONDECYT  Grants 1070306 (SdC), 1090613
(RH and SdC ). Also it was supported by Pontificia Universidad
Catolica de Valparaiso  through grants 123.787-2007 (SdC) and 123.
703-2009 (RH). One of us (E.I.G) would like to thank the
astrophysics and cosmology group at the Pontificia Universidad
Catolica de Valparaiso for hospitality, P. L. is supported by
FONDECYT grant N$^0$ 11090410 and by Direcci\'on de
Investigaci\'on de la Universidad del B\'{\i}o-B\'{\i}o through
Grant N$^0$ 096207 1/R.

\break

\end{document}